\magnification\magstep1

\centerline{\bf New strategy for suppressing decoherence in
quantum computation}

\centerline{Miroljub Dugi\' c}

\centerline{Department of Physics, Faculty of Science, 
Kragujevac, Yugoslavia}

\bigskip

{\bf Abstract:} Controlable strong interaction of the
qubit's bath with an external system (i.e. with the
bath's environment) allows for choosing the conditions
under which the decoherence of the qubit's states can
be substantially decreased (in a certain limit: 
completely avoided). By "substantially decreased" we
mean that the correlations which involve the bath's
states prove negligible, while the correlations
between the qubit's and the environment's states
can be made ineffective during a comparatively long
time interval. So, effectively, one may choose the
conditions under which, for sufficiently long time
interval, the initial state of "qubit + bath" remains
unchanged, thus removing any kind of the errors.
The method has been successfully employed in the
(simplified) model of the
solid-state-nuclear quantum computer (proposed by Kane).

\bigskip

{\bf 1. Introduction}

\bigskip

The issue of decoherence in quantum computation is
one of the central subjects in both fundamentals and
practical realizability of the quantum computers.

Here we propose a new strategy (method) for 
substantial suppression of decoherence in quantum 
computers for arbitrary type of the errors. As opposite
to the existing approaches/methods [1], our strategy is
simple both conceptualy and mathematicaly.

Physicaly, the {\it central idea} of our approach
relies on the following assumptions: {\bf (i)} 
extension of the composite system "qubit + bath (Q+B)"
with the bath's environment, so obtaining the new
composite system "qubit + bath + (bath's) environment
(Q+B+E)", and {\bf (ii)} assumption of existence of
{\it robust states} of the bath which can be selected
(and kept (approximately) unchanged) by {\it
controlable strong} (quantum-measurement-like) 
interaction of the bath with its environment.

Given the points (assumptions) (i) and (ii), the
standard {\it stationary perturbation theory}
points out the next possibiity: with a proper
choice of the initial state of the bath, one may
obtain {\it substantial suppression of decoherence}:
the correlations involving the
bath's states are "(virtually) arbitrarily small", while for
comparatively long time interval the initial
state of Q+B remains unchanged, thus removing
any kind of the errors!

We give a brief account of the issues naturally
appearing in this context, appointing (without
details) that the method works in the 
(simplified) model of the solid-state-nuclear
quantum computer.

\bigskip

{\bf 2. General background}

We employ the standard assumptions of the general
decoherence theory [2,3]: (a) initially, the Q+B+E
system is decoupled, and (b) the dominant terms are
the interaction-Hamiltonians, $\hat H_{QB} +
\hat H_{BE}$. Then we apply the standard algebra [2,3],
based on the direct use of the unitary time-evolution
of the composite system Q+B+E.

One should remind that the standard models of the theory
of decoherence (including the considerations in quantum 
computation) deal with the composite system Q+B, where 
the interaction $\hat H_{QB}$ usually is of the type
(of the separable kind [3]):

$$\hat H_{QB} = c \sum_{p,q} \gamma_{pq} \hat P_{Qp}
\otimes \hat \Pi_{Bq}, \eqno (1)$$

\noindent
where $\hat P_{Qp}$ and $\hat \Pi_{Bq}$ represent the
projectors onto the corresponding subspaces of the
Hilbert state-spaces of Q, and of B, respectively;
$c$ represents the coupling constant.

Then the decoherence is loosely (but sufficiently)
presented by the (time) decrease of the
off-diagonal elements of the Q's
"density matrix":

$$\rho_{Qpp'} = C_p C^{\ast}_{p'}
z_{pp'}(t), \eqno (2a)$$

\noindent
where the "correlation amplitude" $z_{pp'}(t)$" [2]:

$$z_{pp'}(t) = \sum_q p_q \exp(-\imath c t 
(\gamma_{pq} - \gamma_{p'q})/\hbar), \quad \sum_q p_q = 1;
\eqno (2b)$$

\noindent
notice the dependence on the eigenvalues of $\hat
H_{QB}$.

It is important to note that in the context of the,
so-called, {\it macroscopic considerations} [4],
the states enumerated by the indeces $p, p'$ -
which are the elements of the "pointer basis" [2] -
are considered to be {\it robust under the influence
of the environment}. I.e., most of the basic
assumptions necessary for the "transition from
quantum to classical" [5, 6] presuppose invariance
of the "pointer basis" under the transformations
generated by the interaction Hamiltonian. In 
idealized form it reads:

$$\hat H_{QB} \vert \Psi_p\rangle_Q \otimes \vert
\chi\rangle_B = \vert \Psi_p\rangle_Q \otimes \vert
\chi'\rangle_B, \eqno (3)$$

\noindent
where $\vert \Psi_p\rangle_Q$ is an element of the 
"pointer basis" - i.e., of the set of states
bearing the (semi)classical character [2-4] of an open
quantum system.

\bigskip

{\bf 3. The strategy}

\bigskip

We extend the system Q+B by the bath's environment
(E): thus dealing with the new composite system
Q+B+E.

The methodology distingusihed in Section 2 points 
to the following assumptions: (i) initially, the system
is in uncorrelated state $\vert \Psi\rangle_Q \otimes
\vert 0\rangle_B \otimes \vert \chi\rangle_E$, and
(ii) the unitary-evolution operator $\hat U_{QBE}$
can be written as:

$$\hat U_{QBE} \cong \exp(-\imath t (\hat H_{QB} +
\hat H_{BE})/\hbar). \eqno(4)$$

These {\it standard assumptions} are extended by
the following {\it crucial assumptions}: {\bf
(A)} The interaction Hamiltonian $\hat H_{BE}$:

$$\hat H_{BE} = C \sum_{i,j} \kappa_{ij} 
\hat P_{Bi} \otimes \hat \Pi_{Ej}, \eqno (5)$$

\noindent
(compare to eq.(1)) is the {\it dominant term},
thus making $\hat H_{QB}$ {\it the perturbation}
($C$ is the coupling constant), and {\bf (B)}
The initial state $\vert 0\rangle_B$ can be
chosen such that one may state (cf. eq. (3)):

$$\hat H_{BE} \vert 0\rangle_B \otimes 
\vert \chi\rangle_E = \vert 0\rangle_B \otimes 
\vert \chi'\rangle_E. \eqno (6)$$

Then one may employ the standard stationary
perturbation theory [7].

\bigskip

{\bf 3.1 The perturbation theory employed}

\medskip

The basic idea of the perturbation theory [7] 
is presented by the following expressions::

$$\hat H_{BE} \vert \Phi_n\rangle_{QBE} = 
E^{(0)}_n \vert \Phi_n\rangle_{QBE}, \eqno(7) $$
$$(\hat H_{BE} + \hat H_{QB}) \vert \Psi_n\rangle_{QBE}
= E_n \vert \Psi_n\rangle_{QBE}, \eqno (8)$$

\noindent
where in the limit $c \to 0$ one has: $\vert \Psi_n\rangle_{QBE}
\to \vert \Phi_n\rangle_{QBE}$, and $E_n \to E^{(0)}_n$.
(Remind that: the coupling constant $c$ is given and the
above limit should not be literaly understood (it is here for the
formal completeness of the considerations); what we shall
further need is the ratio of the coupling constants $c/C$,
where the limit $c/C \to 0$ is legitimate.)

As it directly follows from eq.(5), the states $\vert \Phi
\rangle_{QBE}$ can be chosen as $\vert pij\rangle \equiv
\vert p\rangle_Q \otimes \vert i\rangle_B \otimes
\vert j\rangle_B$. Then, in accordance with (7) and (8)
one may write for the normalized eigenstates:

$$\vert \Psi_n\rangle_{QBE} \equiv \vert 
\Psi_{pij}\rangle_{QBE} = 
(1 - \epsilon_{pij}^2)^{1/2}\vert pij\rangle + 
\epsilon_{pij} \vert \chi_{pij}\rangle, \eqno (9)$$

\noindent
where $\langle pij\vert \chi_{pij}\rangle = 0$ and
$\langle \chi_{pij}\vert \chi_{pij}\rangle =1$, and

$$E_n \equiv E_{pij} = E_{pij}^{(0)} +
\lambda_{pij} \equiv C \kappa_{ij} + \lambda_{pij}.
\eqno (10)$$

Notice: the corrections of the $\hat H_{BE}$'s
eigenstates and eigenvalues are $\epsilon_{pij} 
\vert \chi_{pij}\rangle$ and $\lambda_{pij}$, respectively,
while eqs.(9, 10) are {\it exact}!

Then from eq. (4), (9), (10), one directly obtains:

$$\hat U_{QBE} \cong \hat U_1 + \hat U_2, \eqno (11a)$$

\noindent
where

$$\hat U_1 = \sum_{(pij)} \exp(-\imath t E_{pij}/\hbar)
(1 - \epsilon_{pij}^2) \vert pij\rangle \langle pij\vert,
\eqno(11b)$$

\noindent
where the sum runs over the different combinations of the
indeces "p, i, j". Bearing in mind that $\Vert \hat U_1 +
\hat U_2\Vert = 1$, it is a matter of straightforward 
algebra to prove that:

$$1 = n_1 + n_2, \quad n_1 = \langle \Psi\vert \hat U_1^{\dag} 
\hat U_1\vert \Psi\rangle, \eqno (12)$$

\noindent
and

$$n_1 \ge (1 - \epsilon_{max}^2)^2, \eqno (12b)$$

\noindent
where $\epsilon_{max}$ is the maximal value of $\epsilon$s
defined by eq.(9).

So, applying $\hat U_{QBE}$ onto the initial state
$\vert \Psi\rangle_Q \otimes \vert 0\rangle_B \otimes
\vert \chi\rangle_E$, one obtains:

$$\hat U_{QBE} \vert \Psi\rangle_Q \otimes \vert 0\rangle_B 
\otimes \vert \chi\rangle_E \cong \sum_{(pij)} C_p \alpha_i
\beta_j \vert pij\rangle + O(\epsilon_{max}), \eqno (13)$$

\noindent
where $C_p = \langle p\vert \Psi\rangle$,
$\alpha_i = \langle i\vert 0\rangle$,  
$\beta_j = \langle j\vert \chi\rangle$, and the bases
$\{\vert i\rangle_B\}$ and $\{\vert j\rangle_E\}$ 
diagonalize [3] $\hat H_{BE}$.

With the choice (cf. above point (ii)) presented by
eq.(6), $\alpha_i = \delta_{ii_{\circ}}$, one obtains:

$$\hat U_{QBE} \vert \Psi\rangle_Q \otimes \vert 0\rangle_B 
\otimes \vert \chi\rangle_E \cong \sum_{(pj)} C_p \beta_j
\exp(-\imath t E_{p0j}/\hbar) \vert p0j\rangle
+ O(\epsilon_{max}). \eqno(14)$$

Now, since

$$E_{p0j} = C \kappa_{0j} + \lambda_{p0j}, \eqno(15)$$

\noindent
and

$$\vert \lambda_{p0j}\vert \le (1- \epsilon_{p0j}^2)^{-1/2} 
\vert \langle p0j\vert 
\hat H_{QB}\vert p0j\rangle\vert +$$
$$+ \vert \epsilon_{p0j}\vert 
(1 - \epsilon_{p0j}^2)^{-1} \vert \langle p0j\vert \hat H_{QB}
\vert \chi_{p0j}\rangle\vert \eqno(16)$$

\noindent
one obtains the {\it main result of this paper}:

if one may choose $\vert 0\rangle_B$ so as the maximal
value ($\lambda_{max}$) of $\vert\lambda_{poj}\vert$s can be very small,
then one may speak of the {\it substantial suppression
of decoherence}.

Actually, since it can be estimated that 
$\epsilon_{max} \sim c/C$, one may say that 
the correlations between Q and B ({\it and} E) can be
considered arbitrarily small, the occurrence of the errors
which come {\it together with the change of B's state} represent
substantially rare events: the total probability of these errors
not exceding the order of $(c/C)^2$, where $C$ {\it is virtually 
arbitrary}.
Now, for $\lambda_{max}$ {\it very small}, one may write:

$$\hat U_{QBE} \vert \Psi\rangle_Q \otimes \vert 0\rangle_B 
\otimes \vert \chi\rangle_E \cong \sum_{(pj)} C_p \vert p\rangle_Q$$
$$\otimes \vert 0\rangle_B \otimes \sum_j \beta_j
\exp(-\imath t C \kappa_{0j}/\hbar)\vert j\rangle_E$$
$$\equiv \vert \Psi\rangle_Q \otimes \vert 0\rangle_B 
\otimes \sum_j \beta_j \exp(-\imath C t \kappa_{0j}/\hbar)
\vert j\rangle_E. \eqno(17)$$

\noindent
{\it for at least the time interval} $\tau$:

$$\tau \sim (\lambda_{max}/\hbar)^{-1}. \eqno (18)$$

I.e. during this time interval, the correlations between the
states of Q and E do not become effective. (Notice: the situation 
with this regard is even much better, for the correlations are 
"driven" by the second term on the r.h.s. of eq.(16)!) 

\bigskip

{\bf 3.2 Physical interpretation}

\medskip

The method directly points to the next {\it protocol for
avoiding the errors in quantum computation}:

\item{}
{\it First, an effective, quantum-measurement-like action of
the environment on the bath should be performed, which serves 
for preparing a robust initial state $\vert 0\rangle_B$. Once
such a state is obtained, the interaction between B and E should 
be strenghtened and prolonged in time, for at least the
interval $\tau$. If the interaction is sufficiently strong
($c/C \ll 1$), then, for the proper choice of the initial state
of the bath, the above algebra guaranties that for the
interval of the order of $\tau$ (eq.(18)), the initial state
of Q+B would appear unchanged, which is sufficient for
preparing the computations in the time intervals much 
shorter than $\tau$. The same procedure should be repeated
for each calculation step.}

Being virtually arbitrary, $\epsilon_{max}$ and $\lambda_{max}$
allow for substantial suppression of decoherence of the
qubit's states, all the effects of decoherence referring
to substantially rare events ($c/C \ll 1$), or falling far beyond the 
interval $\tau$ ($\vert 0\rangle_B$ such that $\lambda_{max} \ll 1$).

Needless to say, in the limits $\epsilon_{max} \to 0$ and 
$\lambda_{max} \to 0$, the r.h.s. of eq.(17) {\it is
exact}! 

\bigskip

{\bf 4. Discussion}

\bigskip

The strategy can be elaborated along the following lines:

\item{(1)}
Generalizations concerning
the interactions of the
qubits themselves,
likewise the self-Hamiltonians of
Q, B, and E.

\item{(2)}
Existence of unknown part of the bath; i.e., that E
interacts with $B_1$, but not with $B_2$ (the real 
bath then would be $B = B_1 + B_2$)

\item{(3)}
Extension of the (well known) bath $B$, if it is not
sufficiently "macroscopic" as to provide us with the
robust states

\item{(4)}
Existence of the common bath for all the qubits 

\item{(5)}
Avoiding the interaction of Q and E in the realistic
situations

\item{(6)}
Considerations of the "classical environment(E)" [8],
inculding the "mixed" initial states of B and E.

As regards the points (1)-(4), the results are
encouraging. 

The work is in progress as regards the points (5,6).

The {\it method has been successfully employed in the
simplified model of the solid-state-nuclear} quantum computer
proposed by Kane[9]. The details will be presented 
elsewhere.

\bigskip

{\bf 5. Conclusion}

\bigskip

The decoherence in quantum computers can be, at
least in principle, suppressed. The idea is to 
properly, strongly "press" the qubit(s)'s bath, and to
produce: the stochastic change of the initial
state of Q+B represents an improbable event
(in the limit $c/C \to 0$ it is stochastically
impossible), and for sufficiently long time
interval this state remains unchanged. So,
one may say that we use decoherence (on B), to
combat decoherence (on Q).

In practical realizations one should try to
choose the interaction of B and E which should be
very strong ($C \gg c$), and such the initial
state of B so as to one may state
$\vert \langle p0j\vert \hat H_{QB}\vert p0j
\rangle \vert \ll 1$. {\bf These choices are really
a matter of the particular model !} The 
preliminarysuccess with the
solid-state-nuclear computer is encouraging.

After the interval $\tau$, the correlations between Q 
and E become effective, leading to decoherence
presented by:

$$\rho_{Qpp'} = C_p C^{\ast}_{p'}
z_{pp'}(t),$$

\noindent
where (compare to eq.(2a)):

$$z_{pp'}(t) = \sum_j \vert \beta_j \vert^2 \exp(-\imath t 
(\lambda_{p0j} - \lambda_{p'0j})/\hbar)$$

\bigskip

\item{[1]}
P. Zanardi and M. Rasetti, Phys. Rev. Lett. {\bf 79} (1997)
3306; A.M. Steane, Phys. Rev. Lett. {\bf 77} (1996) 793;
L. Viola and S. Lloyd, Phys. Rev. A{\bf 58} (1998) 2733;
and references therein

\item{[2]}
W.H. Zurek, Phys. Rev. D {\bf 26}, 1862 (1982)

\item{[3]}
M. Dugi\' c, Physica Scripta {\bf 53}, 9 (1996);
Physica Scripta {\bf 56}, 560 (1997) 

\item{[4]}
W.H. Zurek, Prog. Theor. Phys.
{\bf 89}, 281 (1993) 

\item{[5]}
W.H. Zurek, Phys. Today, October 1991, p. 26

\item{[6]}
R. Omn\' es,  1994.
"The Interpretation of Quantum Mechanics", Princeton University
Press, Princeton, New Jersey

\item{[7]}
A. Messiah, "Quantum Mechanics", North-Holland Publishing
Co., Amsterdam, 1961 (Vol. II, Chap. XVI.17); T. Kato,
Prog. Theor. Phys. {\bf 4}, 154 (1949); C. Bloch, Nucl. Phys.
{\bf 6}, 329 (1958)

\item{[8]}
P. Grigolini, "Quantum mechanical irreversibility and
measurement", World Scientific, Singapore, 1993

\item{[9]}
B.E. Kane, Nature {\bf 323}, 153 (1998)

\end